\documentclass[12pt]{article}
\usepackage{amsfonts}
\usepackage{amssymb}
\usepackage{graphics,psboxit,amsmath}
\usepackage{subfigure}
\usepackage{graphicx}
\usepackage{verbatim}
\usepackage{epic}

\def\hybrid{\topmargin 10pt \oddsidemargin 0pt 
        \headheight 0pt \headsep 0pt
        \textwidth 16,0cm 
        \textheight 22,5cm 
        \marginparwidth .875in
        \parskip 5pt plus 1pt \jot = 1.5ex}

\hybrid

\def\baselinestretch{1.2}

\catcode`\@=11

\def\marginnote#1{}
\newcount\hour
\newcount\minute
\newtoks\amorpm
\hour=\time\divide\hour by60
\minute=\time{\multiply\hour by60 \global\advance\minute by-\hour}
\edef\standardtime{{\ifnum\hour<12 \global\amorpm={am}%
        \else\global\amorpm={pm}\advance\hour by-12 \fi
        \ifnum\hour=0 \hour=12 \fi
        \number\hour:\ifnum\minute<10 0\fi\number\minute\the\amorpm}}
\edef\militarytime{\number\hour:\ifnum\minute<10 0\fi\number\minute}

\def\draftlabel#1{{\@bsphack\if@filesw {\let\thepage\relax
   \xdef\@gtempa{\write\@auxout{\string
      \newlabel{#1}{{\@currentlabel}{\thepage}}}}}\@gtempa
   \if@nobreak \ifvmode\nobreak\fi\fi\fi\@esphack}
        \gdef\@eqnlabel{#1}}
\def\@eqnlabel{}
\def\@vacuum{}
\def\draftmarginnote#1{\marginpar{\raggedright\scriptsize\tt#1}}

\def\draft{\oddsidemargin -.5truein
        \def\@oddfoot{\sl preliminary draft \hfil
        \rm\thepage\hfil\sl\today\quad\militarytime}
        \let\@evenfoot\@oddfoot \overfullrule 3pt
        \let\label=\draftlabel
        \let\marginnote=\draftmarginnote
   \def\@eqnnum{(\theequation)\rlap{\kern\marginparsep\tt\@eqnlabel}%
\global\let\@eqnlabel\@vacuum} }

\def\draft2{
        \def\@oddfoot{\sl preliminary draft \hfil
        \rm\thepage\hfil\sl\today\quad\militarytime}
        \let\@evenfoot\@oddfoot \overfullrule 3pt
        \let\label=\draftlabel
        \let\marginnote=\draftmarginnote
   \def\@eqnnum{(\theequation)\rlap{\kern\marginparsep\tt\@eqnlabel}%
\global\let\@eqnlabel\@vacuum} }

\def\preprint{\twocolumn\sloppy\flushbottom\parindent 2em
        \leftmargini 2em\leftmarginv .5em\leftmarginvi .5em
        \oddsidemargin -.5in \evensidemargin -.5in
        \columnsep .4in \footheight 0pt
        \textwidth 10.in \topmargin -.4in
        \headheight 12pt \topskip .4in
        \textheight 6.9in \footskip 0pt
        \def\@oddhead{\thepage\hfil\addtocounter{page}{1}\thepage}
        \let\@evenhead\@oddhead \def\@oddfoot{} \def\@evenfoot{} }

\def\numberbysection{\@addtoreset{equation}{section}
        \def\theequation{\thesection.\arabic{equation}}}

\def\underline#1{\relax\ifmmode\@@underline#1\else
        $\@@underline{\hbox{#1}}$\relax\fi}

\def\titlepage{\@restonecolfalse\if@twocolumn\@restonecoltrue\onecolumn
     \else \newpage \fi \thispagestyle{empty}\c@page\z@
        \def\thefootnote{\fnsymbol{footnote}} }

\def\endtitlepage{\if@restonecol\twocolumn \else \newpage \fi
        \def\thefootnote{\arabic{footnote}}
        \setcounter{footnote}{0}} 

\catcode`@=12
\relax

\def\figcap{\section*{Figure Captions\markboth
        {FIGURECAPTIONS}{FIGURECAPTIONS}}\list
        {Figure \arabic{enumi}:\hfill}{\settowidth\labelwidth{Figure
999:}
        \leftmargin\labelwidth
        \advance\leftmargin\labelsep\usecounter{enumi}}}
 \relax
\def\tablecap{\section*{Table Captions\markboth
        {TABLECAPTIONS}{TABLECAPTIONS}}\list
        {Table \arabic{enumi}:\hfill}{\settowidth\labelwidth{Table
999:}
        \leftmargin\labelwidth
        \advance\leftmargin\labelsep\usecounter{enumi}}}
 \relax
\def\reflist{\section*{References\markboth
        {REFLIST}{REFLIST}}\list
        {[\arabic{enumi}]\hfill}{\settowidth\labelwidth{[999]}
        \leftmargin\labelwidth
        \advance\leftmargin\labelsep\usecounter{enumi}}}
 \relax
\makeatletter
\newcounter{pubctr}
\def\publist{\@ifnextchar[{\@publist}{\@@publist}}
\def\@publist[#1]{\list
        {[\arabic{pubctr}]\hfill}{\settowidth\labelwidth{[999]}
        \leftmargin\labelwidth
        \advance\leftmargin\labelsep
        \@nmbrlisttrue\def\@listctr{pubctr}
        \setcounter{pubctr}{#1}\addtocounter{pubctr}{-1}}}
\def\@@publist{\list
        {[\arabic{pubctr}]\hfill}{\settowidth\labelwidth{[999]}
        \leftmargin\labelwidth
        \advance\leftmargin\labelsep
        \@nmbrlisttrue\def\@listctr{pubctr}}}
 \relax
\makeatother

\def\ba{\begin{equation}}
\def\ea{\end{equation}}

\def\no{\noindent}

\def\IR{\relax{\rm I\kern-.18em R}}

\begin{document}


\renewcommand{\theequation}{\thesection.\arabic{equation}}
\csname @addtoreset\endcsname{equation}{section}

\newcommand{\eqn}[1]{(\ref{#1})}
\newcommand{\be}{\begin{eqnarray}}
\newcommand{\ee}{\end{eqnarray}}
\newcommand{\non}{\nonumber}
\begin{titlepage}
\strut\hfill
\vskip 1.3cm
\begin{center}


{\bf {\Large On boundary super algebras}}

\vspace{0.5in}

{\large \bf Anastasia Doikou}

 \vskip 0.02in

{\footnotesize University of Patras, Department of Engineering Sciences,\\
GR-26500, Patras, Greece}\\[2mm]
\noindent
{\footnotesize {\tt E-mail: adoikou$@$upatras.gr}}\\
\end{center}

\vskip 1.0in

\begin{abstract}

We examine the symmetry breaking of super algebras due to the presence of appropriate integrable boundary conditions.
We investigate the boundary breaking symmetry associated to both reflection algebras and twisted super Yangians.
We extract the generators of the resulting boundary symmetry as well as we provide explicit expressions of the
associated Casimir operators.

\end{abstract}

\vfill
\baselineskip=16pt

\end{titlepage}

\tableofcontents

\def\baselinestretch{1.2}
\baselineskip 20 pt
\no

\section{Introduction}

Symmetry breaking processes are of the most fundamental concepts in physics.
It was shown in a series of earlier works (see e.g.
\cite{done}--\cite{crampedoikou}), within the context of
quantum integrability, that the presence of suitable boundary conditions may break
a symmetry down without spoiling the integrability
of the system.
It was also shown \cite{doikousfetsos} that the breaking symmetry mechanism due to the presence of integrable boundary
conditions
may be utilized to provide certain centrally extended algebras.
Here, we shall investigate in detail the resulting boundary
symmetries
in the context of quantum integrable models associated to various super-algebras.

More precisely, the main aim of the present work is the study of the algebraic structures underlying quantum
integrable
systems
associated to certain super-algebras,
${\cal Y}(gl({\mathrm m}|{\mathrm n}))$ and
$U_q(gl({\mathrm m}|{\mathrm n})$, once non-trivial boundary conditions are implemented. In this investigation
we shall focus
on the relevant algebraic
content, and our primary objectives will be the study of the related exact symmetries as well as the
construction of the relevant Casimir operators.
The
existence of centrally extended super algebras emerging from these boundary algebras is one of the main motivations
for the present
investigation, and will be discussed in full detail in a forthcoming work given that is a separate significant topic.

We shall focus here on the super Yangian  ${\cal Y}(gl({\mathrm m}|{\mathrm n}))$ and its $q$-deformed
counterpart the
$U_q(gl({\mathrm m}|{\mathrm n}))$ algebra.
It is necessary to first introduce some useful notation associated to super algebras. Consider the ${\mathrm m}+{\mathrm n}$
dimensional
column vectors $\hat e_i$, with 1 at position $i$ and zero everywhere else, and the
$({\mathrm m}+{\mathrm n}) \times ({\mathrm m}+{\mathrm n})$ $e_{ij}$ matrices: $(e_{ij})_{kl} = \delta_{ik}\ \delta_{jl}$.
Then define the grades:
\be
[\hat e_i] = [i], ~~~~~[e_{ij}]= [i]+[j].
\ee
The tensor product is also graded as:
\be
(A_{ij}\otimes A_{kl}) (A_{mn}\otimes A_{pq}) = (-1)^{([k]+[l])([m]+[n])} A_{ij} A_{mn} \otimes A_{kl} A_{pq}.
\ee
Define also the transposition
\be
A^T = \sum_{i,j=1}^{{\mathrm m}+{\mathrm n}} (-1)^{[i][j]+[j]} e_{j i} \otimes A_{ij},
~~~~~~A = \sum_{i,j=1}^{{\mathrm m}+{\mathrm n}} e_{ij} \otimes A_{ij},
\ee
and the super-trace as:
\be
str A = \sum_i(-1)^{[i]} A_{ii}.
\ee
It will also be convenient for our purposes here to define the super-transposition as:
\be
A^t = V^{-1}\ A^T\ V
\ee
where the matrix $V$ will be defined later in the text when appropriate.
Also it is convenient for what follows to introduce the distinguished and symmetric grading,
corresponding apparently
to the distinguished and symmetric Dynkin diagrams. In the distinguished grading we define:
\begin{equation}
[i]=\left\{ \begin{array}{ll}  0\,, & 1 \leq i \leq {\mathrm m}\,,\\
			       1\,, & {\mathrm m}+1\leq i \leq {\mathrm m}+{\mathrm n}\,.
  \end{array}\right. \label{dist}
\end{equation}
In the $gl({\mathrm m}|2{\mathrm k})$ we also define the symmetric grading as:
\begin{equation}
[i]=\left\{ \begin{array}{ll}  0\,, & 1 \leq i \leq {\mathrm k}\,, ~~~~~{\mathrm m}+{\mathrm k} +1\leq i \leq
{\mathrm m}+{2\mathrm k}\\
			       1\,, & {\mathrm k}+1\leq i \leq {\mathrm m}+{\mathrm k}\,.
  \end{array}\right. \label{symm}
\end{equation}

\section{The super Yangian ${\cal Y}(gl({\mathrm m}|{\mathrm n}))$ }

Let us first introduce the basic algebraic objects associated to the Yangian
${\cal Y}(gl({\mathrm m}|{\mathrm n}))$.
The $R$ matrix solution of the Yang-Baxter equation \cite{baxter} associated to
${\cal Y}(gl({\mathrm m}|{\mathrm n}))$
is \cite{kulishsuper, nazarov, yang, saleur}:
\be
R(\lambda) = \lambda + i P \label{syang}
\ee
where $P$ is the super-permutation operator defined as:
\be
P = \sum_{i,j} (-1)^{[j]} e_{ij} \otimes e_{ji}.
\ee
Also define
\be
&& \bar R_{12}(\lambda) := R_{12}^{t_1}(\lambda -i\rho),
~~~~\bar R_{21}(\lambda) := R_{12}^{t_2}(-\lambda - i\rho)
~~~~\mbox{and} ~~~~
\bar R_{12}(\lambda) = \bar R_{21}(\lambda)
\non\\
&& \bar \lambda = -\lambda - i\rho, ~~~~~\mbox{and}
~~~~\rho = {{\mathrm n} - {\mathrm m} \over 2}. \label{bar}
\ee
The $\bar R$ matrix may be written as
\be
\bar R_{12}(\lambda)= \bar \lambda + iQ_{12}
\ee
where $Q$ is a projector satisfying
\be
Q^{2} = 2\rho Q, ~~~~~P\ Q =Q\ P = Q.
\ee
Consider also the $L$-operator expressed as
\be
L(\lambda) = \lambda + i {\mathbb P},~~~~~{\mathbb P} = \sum_{a,b} e_{ab} \otimes {\mathbb P}_{ab}
\ee
with ${\mathbb P}_{ab} \in gl(m|n)$. $L$ is a solution of the equation:
\be
R_{12}(\lambda_1-\lambda_2)\ L_1(\lambda_1)\ L_2(\lambda_2)= L_2(\lambda_2)\ L_1(\lambda_1)\
R_{12}(\lambda_1-\lambda_2)
\label{rtt}
\ee
with $R$ being the matrix above (\ref{syang}). The algebra defined by (\ref{rtt}) is equipped with a co product:
let $L(\lambda) = \sum_{i, j} e_{ij} \otimes l_{ij}(\lambda)$
\be
\Delta(L(\lambda)) = L_{02}(\lambda)\ L_{01}(\lambda)\ \Rightarrow\
\Delta(l_{il}(\lambda)) = \sum_j l_{jl}(\lambda)\otimes l_{ij}(\lambda). \label{co1}
\ee
Define also the opposite co product. Let $\Pi$ be the `shift operator'
$\Pi:\ V_1 \otimes V_2 \hookrightarrow V_2 \otimes V_1$
\be
\Delta' = \Pi \circ \Delta \label{op}
\ee
in particular
\be
\Delta'(L(\lambda)) = L_{01}(\lambda) L_{02}(\lambda)\
\Rightarrow\  \Delta'(l_{il}(\lambda)) = \sum_j (-1)^{([i]+[j])([j]+[l])}
l_{ij}(\lambda) \otimes l_{jl}(\lambda).
\label{co2}
\ee
The $L$ co-products are derived by iteration as:
\be
\Delta^{(L)} = (\mbox{id} \otimes \Delta^{(L-1)}) \Delta, ~~~~~\Delta^{'(L)} = (\mbox{id} \otimes \Delta^{(L-1)})
\Delta'. \label{iter}
\ee
Let us now define the super commutator as
\be
\Big [ A,\ B \Big \} = A B - (-1)^{[A][B]} A B.
\ee
It is easy to show from (\ref{rtt}) that ${\mathbb P}_{ab}$ satisfy the $gl({\mathrm m}| {\mathrm n})$ algebra,
which reads as
\be
&& \Big [{\mathbb P}_{ij},\ {\mathbb P}_{kl} \Big \}=0, ~~~~k \neq j, ~~~~i \neq l \non\\
&& \Big [{\mathbb P}_{ij},\ {\mathbb P}_{ki} \Big \} = (-1)^{[i]} {\mathbb P}_{kj}, ~~~~~k\neq j
\non\\
&& \Big [{\mathbb P}_{ij},\ {\mathbb P}_{jl} \Big \} = -(-1)^{[i]([j]+[l]) +[j][l]} {\mathbb P}_{il} ~~~~~i \neq l \non\\
&& \Big [{\mathbb P}_{ij},\ {\mathbb P}_{ji} \Big \} = (-1)^{[i]} ({\mathbb P}_{jj}-{\mathbb P}_{ii}).
\ee

\subsection{The reflection algebra}

This section serves mostly as a warm up, although some alternative proofs for the symmetries are provided,
and explicit
expressions of quadratic Casimir operators are also given.
Consider now the situation of a boundary integrable system described by the reflection equation
\cite{cherednik, sklyanin}, which also provides the exchange relations of the underlying algebra, i.e. the
reflection algebra
\be
R_{12}(\lambda_{1} -\lambda_{2})\ {\mathbb K}_{1}(\lambda_{1})\ R_{21}(\lambda_{1}
+\lambda_{2})\ {\mathbb K}_{2}(\lambda_{2})=
{\mathbb K}_{2}(\lambda_{2})\ R_{12}(\lambda_{1} +\lambda_{2})\
 {\mathbb K}_{1}(\lambda_{1})\ R_{21}(\lambda_{1} -\lambda_{2})\ ,
\label{re}
\ee
As shown in \cite{sklyanin} a tensorial type representation of the reflection algebra is given by:
\be
{\mathbb T}(\lambda)= T(\lambda)\ K(\lambda)\ \hat T(\lambda)
\ee
where we define
\be
\hat T(\lambda) = T^{-1}(-\lambda), ~~\mbox{and}~~
T(\lambda) = \Delta^{(N+1)}(L)= L_{0N}(\lambda) \ldots L_{01}(\lambda)
\ee
$K$ is a $c$-number solution of the reflection equation.

The associated transfer matrix is defined as
\be
t(\lambda) = str\{K^+(\lambda)\ {\mathbb T}(\lambda) \} \label{opentransfer}
\ee
$K^+$ is also a solution of the reflection equation, and
\be
\Big [t(\lambda),\  t(\lambda') \Big ] =0.
\ee

In the special case where $K^+ = K ={\mathbb I}$ the transfer matrix
enjoys the full $gl(m|n)$ symmetry. Here we shall provide an explicit proof based on
the linear relations satisfied by the algebra co products and the matrix ${\mathbb T}$.
The proof of this statement goes as follows:
let us first recall the co-product of the $gl(m|n)$ elements
\be
\Delta({\mathbb P}_{ab}) = {\mathbb I} \otimes {\mathbb P}_{ab} + {\mathbb P}_{ab} \otimes {\mathbb I}.
\ee
Define also the following representation $\pi:\ gl(m|n) \hookrightarrow \mbox{End}({\mathbb C}^{n+m})$ such that:
$\pi({\mathbb P}_{ab}) = P_{ab}$.

The $N+1$ co-product satisfies the following commutation relations with the monodromy matrix
\be
(\pi \otimes \mbox{id}^{\otimes N} ) \Delta^{(N+1)}({\mathbb P}_{ab})\ T(\lambda) =
T(\lambda)\ (\pi \otimes \mbox{id}^{\otimes N}) \Delta^{(N+1)}({\mathbb P}_{ab}).
\ee
The later relations may be written in a more straightforward form as:
\be
\Big (P_{ab}\otimes {\mathbb I}  + {\mathbb I} \otimes \Delta^{(N)}({\mathbb P}_{ab})\Big )\ T(\lambda) =
T(\lambda)\ \Big (P_{ab}\otimes {\mathbb I}  + {\mathbb I} \otimes \Delta^{(N)}({\mathbb P}_{ab})\Big ).
\label{simple}
\ee
It is also clear that $T^{-1}(-\lambda)$ also satisfies relations (\ref{simple}), so
for $K={\mathbb I}$ it is quite straightforward to show
that (recall also that $P_{ab} = (-1)^{[b]} e_{ab}$)
\be
\Big ((-1)^{[b]} e_{ab} \otimes {\mathbb I} + {\mathbb I} \otimes \Delta^{(N)}({\mathbb P}_{ab})\Big )\
{\mathbb T}(\lambda) =
{\mathbb T}(\lambda)\ \Big ( (-1)^{[b]} e_{ab}\otimes {\mathbb I}  + {\mathbb I} \otimes \Delta^{(N)}({\mathbb P}_{ab})
\Big ).
\ee
If we now express ${\mathbb T}(\lambda) = \sum_{i,\ j}e_{ij} \otimes {\mathbb T}_{ij}(\lambda) $ then the
latter relations become
\be
&& \sum_j (-1)^{[b]} e_{aj} \otimes {\mathbb T}_{bj}(\lambda) + \sum_{i,j} (-1)^{([a]+[b])([i]+[j])} e_{ij} \otimes
\Delta^{(N)}({\mathbb P}_{ab}){\mathbb T}_{ij}(\lambda)\non\\
&& = \sum_i(-1)^{[b]+([a]+[b])([a] +[i])} e_{ib} \otimes {\mathbb T}_{ia}(\lambda) + \sum_{i, j} e_{ij}
\otimes {\mathbb T}_{ij}(\lambda) \Delta^{(N)}({\mathbb P}_{ab}).
\label{basic}
\ee
We are however interested in the super-trace over the auxiliary space, so we are dealing basically with the diagonal terms of the
above equation,
hence we obtain the following exchange relations:
\be
\Big [ {\mathbb T}_{ii}(\lambda),\ \Delta^{(N)}({\mathbb P}_{ab})\Big ] &=&0, ~~~~~i\neq a,\ i \neq b \non\\
\Big [ {\mathbb T}_{aa}(\lambda),\ \Delta^{(N)}({\mathbb P}_{ab}) \Big ] &=& (-1)^{[b]} {\mathbb T}_{ba}(\lambda),
~~~~
\Big [ {\mathbb T}_{bb}(\lambda),\ \Delta^{(N)}({\mathbb P}_{ab}) \Big ]= -(-1)^{[a]} {\mathbb T}_{ba}(\lambda).
\ee
By taking now the super-trace (we are considering $K^+ = {\mathbb I}$) we have
\be
&& \Big [ \sum_{i}(-1)^{[i]}{\mathbb T}_{ii}(\lambda),\  \Delta^{(N)}({\mathbb P}_{ab}) \Big ] =
\Big [(-1)^{[a]}{\mathbb T}_{aa}(\lambda) +(-1)^{[b]}{\mathbb T}_{bb},\  \Delta^{(N)}({\mathbb P}_{ab}) \Big ]
=\ldots =0 \non\\
&& \Rightarrow \Big [t(\lambda),\  \Delta^{(N)}({\mathbb P}_{ab})\Big ] =0
\ee
and consequently:
\be
\Big [t(\lambda),\  gl( {\mathrm m}|{\mathrm n})\Big ] = 0.
\ee

Recall that here we are focusing on the distinguished Dynkin diagram (\ref{dist}).
Consider now the non trivial situation where the $K$-matrix has the following diagonal form (see also \cite{annecy1}).
\be
K(\lambda) = \mbox{diag} (\underbrace{1, \dots 1}_{m_1},
\underbrace{-1, \ldots -1}_{m_2+n_2}, \underbrace{1, \dots ,1}_{n_1}) \label{kka}
\ee
such that ${\mathrm m}= {\mathrm m}_1 +{\mathrm m}_2,\ {\mathrm n} ={\mathrm n}_1 +{\mathrm n}_2$. In general,
any solution \cite{annecy1} may
be written in the form
$K(\lambda) = i\xi  +\lambda {\cal E}, ~~~{\cal E}^2 = {\mathbb I}$. $\cal E$ may be diagonalized into (\ref{kka})
and that is why we
make this convenient choice for the $K$ matrix (\ref{kka}), we also chose for simplicity  $\xi =0$.

Let us first extract the non-local charges for any generic $K$-matrix of the form:
\be
K(\lambda) = K + {1\over \lambda} \xi_1 +{1\over \lambda^2} \xi_2 + {\cal O}({1 \over \lambda^3}), \label{generic}
\ee
(see also \cite{ragoucysatta} for a brief discussion on the symmetry).
From the asymptotic behavior of the dynamical ${\mathbb T}$ we then obtain:
\be
{\mathbb T}(\lambda \to \infty) \sim K + {i \over \lambda}{\mathbb Q}^{(0)}  -{1 \over \lambda ^2}{\mathbb Q}^{(1)}
 + \ldots
\ee
The first order quantity provides the generators of the remaining boundary symmetry,
\be {\mathbb Q}^{(0)} = \Delta^{(N)}({\mathbb P})K + K \Delta^{(N)}({\mathbb P}) +\xi_1
\ee
and for the special choice of $K$-matrix (\ref{kka}) we conclude
\be
&& {\mathbb Q}^{(0)}=  \sum_{i, j=1}^{{\mathrm m}_1} e_{ij} \otimes \Delta^{(N)}({\mathbb P}_{ij})
+ \sum_{i, j={\mathrm m}+{\mathrm n}_2}^{{\mathrm m} + {\mathrm n}} e_{ij} \otimes \Delta^{(N)}({\mathbb P}_{ij}) -
\sum_{i, j=1{\mathrm m}_1 +1}^{{\mathrm m}+{\mathrm n}_2} e_{ij} \otimes \Delta^{(N)}({\mathbb P}_{ij}) + \non\\
&&  \sum_{i=1}^{{\mathrm m}_1} \sum_{j = {\mathrm m}+{\mathrm n}_2 +1}^{{\mathrm m} + {\mathrm n}}e_{ij} \otimes
\Delta^{(N)}({\mathbb P}_{ij})
+\sum_{i = {\mathrm m}+{\mathrm n}_2 +1}^{{\mathrm m} + {\mathrm n}} \sum_{j=1}^{{\mathrm m}_1} e_{ij} \otimes
\Delta^{(N)}({\mathbb P}_{ij}).
\ee
More precisely, the elements:
\be
&& {\mathbb P}_{ij},\ ~~i,j \in (1,\ {\mathrm m}_1)\cup ({\mathrm m} +{\mathrm n}_2 +1,\
{\mathrm m}+{\mathrm n}) ~~~\mbox{form the}  ~~~~gl({\mathrm m}_1|{\mathrm n}_1) \non\\
&& {\mathbb P}_{ij},\ ~~i,j \in ({\mathrm m}_1+1,\
{\mathrm m}+{\mathrm n}_2)
~~~~~\mbox{form the} ~~~~gl({\mathrm m}_2|{\mathrm n}_2). \label{gens0}
\ee
These are exactly the generators that, in the fundamental representation, commute with the $K$ matrix (\ref{kka}).

That is the $gl({\mathrm m}|{\mathrm n})$ symmetry breaks down to $gl({\mathrm m}_1|{\mathrm n}_1)
\otimes gl({\mathrm m}_2|{\mathrm n}_2)$. Since the $K$-matrix commutes with all the above generators
following the procedure above we can show relations (\ref{simple}) but only with the generators (\ref{gens0}) and finally:
\be
\Big [ t(\lambda),\ gl({\mathrm m}_1|{\mathrm n}_1) \otimes gl({\mathrm m}_2|{\mathrm n}_2)\Big ] =0.
\ee

We shall focus now for simplicity on the `one-particle' representation $N=1$.
The trace of the second order quantity provides the quadratic quadratic Casimir associated to the
$gl({\mathrm m}|{\mathrm n})$ in the case where $K =1$.
When $K$ is of the diagonal form (\ref{kka}) the Casimir (see also \cite{molev0, supercasimir}) is associated
to $gl({\mathrm m}_1|{\mathrm n}_1)
\otimes gl({\mathrm m}_2|{\mathrm n}_2)$.

More specifically, for $K \propto {\mathbb I}$ (set $N=1$):
\be
&& {\mathbb Q}^{(1)} = 2 {\mathbb P}^2\  ~~\mbox{and}~~ \  C = str {\mathbb Q}^{(1)}  = 2
\sum_{i,j=1}^{{\mathrm m} +{\mathrm n}} (-1)^{[j]} {\mathbb P}_{ij}{\mathbb P}_{ji}
\ee
and for $K$ given by the diagonal matrix (\ref{generic})
\be
&& {\mathbb Q}^{(1)} = {\mathbb P}  K {\mathbb P} + K {\mathbb P}^2-i {\mathbb P} \xi_1 -i \xi_1 {\mathbb P} - \xi_2
 \ ~~\mbox{and}~~ \non\\
&& C = str {\mathbb Q}^{(1)}.
\ee
For the special choice of $K$-matrix (\ref{kka}) we have:
\be
C &=& \sum_{i = 1}^{{\mathrm m}_1} \Big ( \sum_{j = 1}^{{\mathrm m}_1}
(-1)^{[j]} {\mathbb P}_{ij}{\mathbb P}_{ji} +
\sum_{i={\mathrm m}+{\mathrm n_2}}^{{\mathrm m} +{\mathrm n}} (-1)^{[j]} {\mathbb P}_{ij}{\mathbb P}_{ji} \Big )
\non\\&
+& \sum_{j = 1}^{{\mathrm m}_1} \Big ( \sum_{i = 1}^{{\mathrm m}_1}
(-1)^{[j]} {\mathbb P}_{ij}{\mathbb P}_{ji} +
\sum_{i = {\mathrm m} + {\mathrm n_2}}^{{\mathrm m} + {\mathrm n}} (-1)^{[j]} {\mathbb P}_{ij}{\mathbb P}_{ji} \Big )
- \sum_{i, j = {\mathrm m}_1 + 1}^{{\mathrm m} + {\mathrm n}_2}(-1)^{[j]} {\mathbb P}_{ij}{\mathbb P}_{ji}.
\ee
Higher Casimir operators may be extracted by considering the higher order terms in the expansion of the
transfer matrix in powers of ${1\over \lambda}$.
More precisely, let us focus on the $N=1$ case and see more precisely how one obtains the higher Casimir operators
from the expansion of the transfer matrix $t(\lambda) = \sum_{k=1}^{2N} {t^{(k-1)} \over \lambda^k}$.
Recall the $N=1$ representation of the reflection algebra
\be
&& {\mathbb T}(\lambda) = L(\lambda)\ k\ \hat L(\lambda) = (1 +{i\over \lambda}{\mathbb P})\ k\
(1 +{i\over \lambda}{\mathbb P} -{1\over \lambda^2}{\mathbb P}^2
-{i \over \lambda^3}{\mathbb P}^3 + {1\over \lambda^4}{\mathbb P}^4 \ldots) \non\\
&& = k + {i\over \lambda}({\mathbb P} k + k {\mathbb P}) -{1 \over \lambda^2}({\mathbb P} k {\mathbb P} +
k{\mathbb P}^2) - {1\over \lambda^2} ({\mathbb P}k {\mathbb P}^2 + k {\mathbb P}^3 ) \ldots \label{exp1}
\ee
where $k$ is diagonal then
\be
t^{(k-1)} \propto \sum_{a, b}({\mathbb P}_{ab}\ k_{bb}\ {\mathbb P}^{k-1}_{ba} + k_{aa}\
{\mathbb P}_{aa}^k  ) \label{tk}
\ee
All $t^{(k)}$ are the higher Casimir quantities and by construction they commute
with each other and they commute as shown earlier with the
exact symmetry of the system.
Depending on the rank of the considered algebra
the expansion of $t(\lambda)$ should truncate at some point; note that expressions (\ref{exp1}), (\ref{tk})
are generic and hold for any $gl({\mathrm m}|{\mathrm n})$. The spectra of all Casimir
operators associated
to a specific algebra may be derived via the Bethe ansatz methodology. In particular, the spectrum
for generic representations of
super symmetric algebras is known (see e.g. \cite{belliardragoucy}).
By appropriately expanding the eigenvalues in powers of ${1\over \lambda}$ we may identify the spectrum of each one of the
relevant Casimir operators.

\subsection{The twisted super Yangian}

Note that we focus here in the $gl({\mathrm m}|2{\mathrm k})$ case and the symmetric Dynkin diagramm (\ref{symm}).
Let us first define some basic notation useful or our purposes here.
Consider the matrix
\be
V= \sum_{i} f_{i}  e_{i \bar i}, ~~~\mbox{where}, ~~~\bar i ={\mathrm m}+2{\mathrm k} - i+1.
\ee
More precisely we shall consider here the following anti-diagonal matrix:
\be
V =\mbox{antidiag} (\underbrace{1, \ldots, 1}_{{\mathrm m} + {\mathrm k}}, \underbrace{-1 ,\dots -1}_{{\mathrm k}}).
\ee

The twisted super Yangian defined by \cite{molev, moras, briotragoucy} (for more details on the physical meaning of
reflection algebra and twisted Yangian
see \cite{doikou0, annecy1}):
\be
R_{12}(\lambda_1 - \lambda_2)\ \bar {\mathbb K}_1(\lambda_1)\ \bar R_{12}(\lambda_1 - \lambda_2)\ \bar
{\mathbb K}_2(\lambda_2)=
\bar {\mathbb K}_2(\lambda_2)\ \bar R_{12}(\lambda_1 - \lambda_2)\ \bar {\mathbb K}_1(\lambda_1)\
R_{12}(\lambda_1 - \lambda_2) \label{re2}
\ee
the matrix $\bar R_{12}$ is defined in (\ref{bar}).

Define also
\be
\hat L_{0n}(\lambda) = L_{0n}^{t_0}(-\lambda - i\rho) \propto 1 + {i \over \lambda} \hat {\mathbb P}_{0n},
~~~\mbox{where}~~~  \hat {\mathbb P}_{0n} = \rho - {\mathbb P}_{0n}^{t_0}.
\ee
Consider now the generic tensorial representation of the twisted super Yangian:
\be
\bar {\mathbb T}(\lambda) = T(\lambda)\ K(\lambda)\ T^{t_0}(-\lambda -i\rho).
\ee
$K$ is a $c$-number solution of the twisted Yangian (\ref{re2}).

As in the previous section express the above tensor representation in powers of $1 \over \lambda $:
\be
\bar {\mathbb T}(\lambda) = 1 + {i\over \lambda}(\bar {\mathbb Q}^{(0)} +  N \rho)-  {1\over \lambda^2}
\bar {\mathbb Q}^{(1)} +\ldots
\ee
where
\be
\bar {\mathbb Q}^{(0)} = \Delta^{(N)}({\mathbb P}) -\Delta^{(N)}({\mathbb P}^{t_0}).
\ee
It is clear that the elements $\bar {\mathbb Q}_{ab}$ form the $osp({\mathrm m}|2{\mathrm n})$ algebra,
and this corresponds essentially to a folding of the
$gl({\mathrm m}|{2\mathrm n})$ to $osp({\mathrm m}| 2 {\mathrm n})$. Such a folding occurs in the
corresponding symmetric Dynkin diagram.
Henceforth, we shall consider the simplest solution $K \propto {\mathbb I}$, although
a full classification is presented in \cite{annecy1}.
Based on the same logic as in the previous paragraph we may extract the corresponding exact symmetry.
We have in this case:
\be
(\pi \otimes \mbox{id}^{\otimes N})\Delta^{(N+1)}(\bar {\mathbb Q}_{ab}^{(0)})\ \bar {\mathbb T}(\lambda) =
\bar {\mathbb T}(\lambda)\
(\pi \otimes \mbox{id}^{\otimes N})\Delta^{(N+1)}(\bar {\mathbb Q}_{ab}^{(0)}),
\ee
which leads to
\be
\Big [t(\lambda),\ osp({\mathrm m}| 2{\mathrm n})\Big ]=0,
\ee
so the exact symmetry of the considered transfer matrix is indeed $osp({\mathrm m}|2{\mathrm n})$.

The quadratic Casimir operator associated to $osp({\mathrm m}| {\mathrm n})$ emerges
from the super-trace of
$\bar {\mathbb Q}^{(1)}$ ($N=1$):
\be
\bar {\mathbb Q}^{(1)} = {\mathbb P}\ \hat {\mathbb P} ~~~~~
C = str \bar{\mathbb Q}^{(1)}  = \sum_{i, j} (-1)^{[j]} {\mathbb P}_{ij} \hat {\mathbb P}_{ji}.
\ee


\section{The $U_q(gl({\mathrm m}|{\mathrm n}))$ algebra}

We come now to the $q$ deformed situation.
The $R$-matrix associated to the $U_q(\widehat{gl({\mathrm m}|{\mathrm n}}))$
algebra is given by the following expressions \cite{perk}:
\be
R(\lambda) = \sum_{i=1}^{{\mathrm m}+{\mathrm n}} a_i(\lambda) e_{ii} \otimes e_{ii}+ b(\lambda)
\sum_{i\neq j=1}^{{\mathrm m}+{\mathrm n}} e_{ii} \otimes e_{jj}
+ \sum_{i\neq j=1}^{m+n} c_{ij}(\lambda) e_{ij} \otimes e_{ji}, \label{rr2}
\ee
where we define
\be
a_j(\lambda) = \sinh(\lambda +i\mu -2 i \mu [j]) ,
~~~~~
b(\lambda) = \sinh \lambda,
~~~~c_{ij}(\lambda) =\sinh(i \mu) e^{sign(j-i)\lambda} (-1)^{[j]}.
\ee

Let us now introduce the super symmetric Lax operator associated to $U_q(\widehat{gl({\mathrm m}|{\mathrm n})})$
\be
L(\lambda) =e^{\lambda} L^+ - e^{-\lambda} L^-
\ee
and $L$ satisfies the fundamental algebraic relation (\ref{rtt}) with the $R$ matrix given in (\ref{rr2}).
The elements $L^{\pm}$ satisfy \cite{tak, fad}
\be
&& R_{12}^{\pm}\ L_1^+\ L^+ = L_2^{+}\ L_1^+\ R_{12}^{\pm}, \non\\
&& R_{12}^{\pm}\ L_1^-\ L^- = L_2^{-}\ L_1^-\ R_{12}^{\pm}, \non\\
&& R_{12}^{\pm}\ L_1^{\pm}\ L_2^{\mp}= L_2^{\mp}\ L_1^{\pm}\ R_{12}^{\pm} \label{+-}
\ee
$L^{\pm}$ are expressed as:
\be
L^+ = \sum_{i\leq j }e_{ij}\otimes l^+_{ij}, ~~~~~~~L^- = \sum_{i\geq j }e_{ij}\otimes l^-_{ij}
\ee
definitions of $l_{ij}^{\pm},\ \hat l^{\pm}_{ij}$ in terms of the $U_q(gl({\mathrm m}|{\mathrm n}))$ algebra
generators in the Chevalley-Serre basis are given in Appendix A.
The equations above (\ref{+-}) provide all the exchange relations
of the $U_q(gl({\mathrm m}|{\mathrm n}))$ algebra. This is in fact the so called FRT realization of
the $U_q(gl({\mathrm m}|{\mathrm n}))$
algebra (see e.g. \cite{tak}).

\subsection{The reflection algebra}

Our main objective here is to extract the exact symmetry of the open transfer matrix associated to
$U_q(gl(\widehat{{\mathrm m}| {\mathrm n}}
))$. We shall
focus in this section on the distinguished Dynkin diagram. The open transfer matrix is given by (\ref{opentransfer}),
and from now on we
consider for our purposes here the left boundary to be the trivial solution
\be
K^+ = M = \sum_{k =1}^{{\mathrm m} + {\mathrm n}} q^{{\mathrm n} + {\mathrm m} -2 k +1} q^{-2[k] + 4
\sum_{i=1}^k [i]} e_{kk}.
\ee
The elements one extracts from the asymptotic expansion of ${\mathbb T}$ by keeping the leading contribution,
${\mathbb T}_{ab}^{\pm}$ are the boundary non-local charges, which form the boundary super algebra with
exchange relations dictated by:
\be
R_{12}^{\pm}\ {\mathbb T}^{\pm}_1\ R_{21}^{\pm}\ {\mathbb T}_2^{\pm}=
{\mathbb T}_2^{\pm}\ R_{21}^{\pm}\ {\mathbb T}^{\pm}_1\ R_{12}^{\pm}.
\ee
In general, it may be shown that the boundary super algebra is an exact symmetry of the double row transfer matrix.
Indeed by introducing the element
\be
\tau^{\pm} = str (e_{ab} {\mathbb T}^{\pm}) = (-1)^{[a]} {\mathbb T}_{ba}^{\pm}
\ee
it is quite straightforward to show along the lines described in \cite{doikoutwin} that
\be
[\tau^{\pm},\ t(\lambda) ] =0  \quad \Rightarrow \quad \big [ {\mathbb
T}_{ab}^{\pm},\ t(\lambda) \big ] =0. \label{symme2}
\ee
Let
\be
&& T^{\pm} = \sum_{ab} e_{ab} \otimes T_{ab}^{\pm}, ~~~~~~\hat T^{\pm} = \sum_{ab} e_{ab} \otimes
\hat T_{ab}^{\pm} \non\\
&& \mbox{and} ~~~~T^{\pm}_{ab} = \Delta^{(N)}(l_{ab}^{\pm}), ~~~~~~\hat T^{\pm}_{ab} =
\Delta^{(N)}(\hat l_{ab}^{\pm}),
\ee
see also Appendix A for definitions of $l^{\pm}_{ij},\ \hat l^{\pm}_{ij}$.

The boundary non-local charges emanating from ${\mathbb T}^{\pm}$ are of the explicit form:
\be
{\mathbb T}_{ad}^{\pm} = \sum_{b,c} (-1)^{([a]+[b])([b]+[d])} T_{ab}^{\pm}\ K_{bc}^+\ \hat T^{\pm}_{cd}. \label{nonlocal}
\ee
It is clear that different choice of $K$-matrix leads to different non-local charges and consequently to
different symmetry.
Also, the quadratic Casimir operators are obtained by
\be
C^{\pm} = str_0(M_0{\mathbb T}_0^{\pm}) = \sum_{a, b, c} (-1)^{[b]} M_{aa} T^{\pm}_{ab}K^{\pm}_{bc}\hat T_{ca}^{\pm}.
\label{qcas}
\ee
Explicit expressions for the quadratic Casimir operators will be given below for particular simple examples.
Note that higher Casimir operators may be extracted from the higher order terms in the
expansion of the transfer matrix in powers of $e^{\pm 2 \lambda}$.

\subsubsection{Diagonal reflection matrices}

We shall distinguish two main cases of diagonal matrices,  and we shall examine the corresponding exact symmetry.
First consider the simplest boundary conditions described by $K \propto {\mathbb I}$, we shall explicitly show that
the associated symmetry
is the $U_q(gl({\mathrm m}|{\mathrm n}))$.
One could just notice that
the resulted ${\mathbb T}^{\pm}$ form essentially the $U_q(gl({\mathrm m}|{\mathrm n}))$,
but this is not quite obvious.
Then  via (\ref{symme2})
one can show the exact symmetry of
the transfer matrix. Let us however here
follow a more straightforward and clean approach regarding the symmetry, that is we shall show that the open transfer
matrix commutes
with each one of the
algebra generators. Our proofs hold for generic non-trivial integrable boundary conditions as will be more transparent subsequently.
In fact, in this and the previous section we prepare somehow the general algebraic setting so that we may
extract the symmetry for generic integrable boundary conditions.

Let us now set the basic algebraic machinery necessary for the proofs that follow.
Our aim now is to show that the double row transfer matrix with trivial boundary conditions that is
$K = {\mathbb I},\ K^{(L)} =M$
enjoys the full $U_q(gl({\mathrm m}|{\mathrm n}))$ symmetry. We shall show in particular
 that the open transfer matrix
commutes with each one of the generators of $U_q(gl({\mathrm m}|{\mathrm n}))$.
Let us outline the proof;
first it is quite easy to show from the form of the co product for $\epsilon_i$ and following
the logic described in the previous section see equation (\ref{simple}) that
\be
[\Delta^{(N)}(\epsilon_i),\ t(\lambda)] =0.
\ee
We may now show the commutation between the transfer matrix and the other elements of the super algebra.
Introduce the representation $\pi:\ U_q(gl({\mathrm m}|{\mathrm n})) \hookrightarrow \mbox{End}({\mathbb C}^{\cal N})$
\be
\pi(e_i) =\varepsilon^i, ~~~~\pi(f_i)= \phi^i, ~~~~\pi(q^{{h_i\over 2}}) = k^{i}. \label{rep1}
\ee
Consider also the following more convenient notation
\be
\Delta^{(N)}(q^{\pm {h_i \over 2}}) \equiv (K_N^i)^{\pm 1}, ~~~~~~\Delta^{(N)}(e_i)\equiv E^i_N,
~~~~~~\Delta^{(N)}(f_i)\equiv F^i_N.
\ee
We shall now make use of the generic relations, which clearly ${\mathbb T}$ satisfies due to the particular
choice of boundary conditions (see also \cite{myhecke}):
\be
(\pi \otimes \mbox{id}^{\otimes N}) \Delta^{'(N+1)}(Y)\ {\mathbb T}(\lambda) =
{\mathbb T}(\lambda)\ (\pi \otimes \mbox{id}^{\otimes N}) \Delta^{'(N+1)}(Y),
~~~~Y\in U_q(gl({\mathrm m}|{\mathrm n})).
\ee

Let us restrict our proof to $e_i$ although the same logic follows for proving the commutation of the transfer
matrix with $f_i$.
In addition to the algebra exchange relations, presented in
Appendix A, we shall need for our proof the following relations:
\be
&& M\ e_i = q^{a_{ii}}\ e_{i}\ M,~~~~~ M\ f_i = q^{-a_{ii}}\ f_{i}\ M\non\\
&& (k_0^{i} K_N^i)^{\pm 1}\ {\mathbb T}_0(\lambda) = {\mathbb T}_0(\lambda)\ (k_0^{i} K_N^i)^{\pm 1}.
\ee
It is convenient to rewrite the relation above for $e_i$ as follows:
\be
&& (k_0^{i} E_N^i+ \varepsilon_0^i (K_N^i)^{-1}){\mathbb T}_0(\lambda) = {\mathbb T}_0(\lambda)(k_0^{i} E_N^i+
\varepsilon_0^i (K_N^i)^{-1})
\ldots \Rightarrow \non\\
&& E^i_N M_0{\mathbb T}_0(\lambda) + q^{{\alpha_{ii} \over 2}}\varepsilon_0^{i} M_0 {\mathbb T}_0(\lambda)
(k_0^i)^{-1}(K^i_N)^{-1} =
(k_0^i)^{-1} M_0 {\mathbb T}_0(\lambda)k_0^i E_N^i + (k_0^i)^{-1} M_0 {\mathbb T}_0(\lambda) \varepsilon_0^i
(K_N^i)^{-1}, \non\\ \label{funda0}
\ee
where the subscript $0$ denotes the auxiliary space, whereas the $N$ co products leave exclusively
on the quantum space,
recall also that
$\varepsilon^i \propto e_{i i +1}$.
By focusing only on the diagonal contributions over the auxiliary space,
after some algebraic manipulations, and bearing in mind (\ref{funda0}),
we end up to the following exchange relations (recall $M$ is diagonal):
\be
&&[E_N^i,\ M_{aa} {\mathbb T}_{aa}(\lambda)]= 0, ~~~a \neq i,\ i+1 \non\\
&&[E_N^i,\ M_{ii} {\mathbb T}_{ii}(\lambda)] = -\rho_i M_{i+1 i+1} {\mathbb T}_{i+1 i}(\lambda)
(K_N^i)^{-1}\non\\
&&[E_N^i,\ M_{i+1 i+1 } {\mathbb T}_{i+1 i+1}(\lambda)] =  (-1)^{[i]+[i+1]} \rho_i M_{i+1 i+1} {\mathbb T}_{i+1 i}
(\lambda) (K_N^i)^{-1}
\ee
$\rho_i$ is a scalar depending on $a_{ii}$ and the grading.
Then based on the relations above we have
\be
[E_N^i,\ \sum_{a=1}^{{\cal N}}(-1)^{[a]} M_{aa} {\mathbb T}_{aa}(\lambda)] =0\ ~\Rightarrow\
~[E_N^i,\ t(\lambda)] =0.
\ee
Similarly one may show the commutativity of the transfer matrix with $F_N^i$, hence
\be
[\Delta^{(N)}(x),\ t(\lambda)] =0, ~~~x \in U_q(gl({\mathrm m}|{\mathrm n})),
\ee
and this concludes our proof on the exact symmetry of the double row transfer matrix for the particular choice
of boundary conditions.
It is clear that the proof above can be easily applied to the usual non super symmetric deformed algebra. We have
not seen,
as far as we know, such an explicit and elegant proof elsewhere not even in the non super symmetric case.
In \cite{myhecke} the proof is explicit, but rather tedious,
whereas in \cite{done1} one has to realize that the emanating non local charges form the $U_q(gl({\cal N}))$, and this is
not quite obvious.
Note that ${\mathbb T}^{\pm}_{ab}$ are
quadratic combinations of the algebra generators, and in fact
the corresponding super-trace provides the associated Casimir, which again gives a hint about the associated exact
symmetry. In the case of trivial boundary conditions ($K \propto {\mathbb I}$)
there is a discussion on the symmetry in \cite{qsupersymmetry}, but is restricted only
in this particular case, whereas
our proof holds for generic non-trivial integrable boundary condition (see also next section).

Let us now give explicit expressions of the quadratic $q$-Casimir for the simplest case (see also \cite{qcasimir}),
that is the $U_q(gl(1|1))$
situation.
In general, the quadratic Casimir operator is given by (\ref{qcas}), but now $K^{\pm}_{ab} = \delta_{ab}$
\be
&& C^+ \propto \Delta^{(N)}( q^{2\epsilon_1}) - \Delta^{(N)}(q^{2\epsilon_2}) -(q-q^{-1})^2
\Delta^{(N)}(q^{{\epsilon_1 +\epsilon_2 \over 2}}) \Delta^{(N)}(f_1) \Delta^{(N)}
(q^{{\epsilon_1 +\epsilon_2 \over 2}})
 \Delta^{(N)}(e_1)
\non\\
&& C^- \propto  \Delta^{(N)}( q^{-\epsilon_1}) - \Delta^{(N)}(-q^{2\epsilon_2}) +(q-q^{-1})^2
\Delta^{(N)}(e_1) \Delta^{(N)}(q^{-{\epsilon_1 +\epsilon_2 \over 2}}) \Delta^{(N)}(f_1) \Delta^{(N)}(q^{-{\epsilon_1
+ \epsilon_2 \over 2}}).
\non\\
\ee

Consider diagonal non-trivial solutions of the reflection equation (see also a brief discussion
on the symmetry in \cite{belliardragoucy}):
\be
K(\lambda) = \mbox{diag}(\underbrace{a(\lambda), \ldots, a(\lambda),}_{\alpha} \underbrace{b(\lambda), \ldots,
b(\lambda)}_{{\cal N}- \alpha}). \label{kdiag}
\ee
In this case it is clear that the the $K$ matrix satisfies (recall (\ref{rep1}))
\be
[\pi(x),\ K(\lambda)] =0\ ~~~~ && x \in U_q(gl({\mathrm m} -\alpha|{\mathrm n})) \otimes U_q(gl(\alpha)),
~~~\mbox{if} ~~\alpha ~~\mbox{bosonic} \non\\
&& x \in  U_q(gl({\mathrm m}|\tilde \alpha)) \otimes U_q(gl({\mathrm n}- \tilde \alpha))~~~\mbox{if} ~~
\alpha = m + \tilde \alpha ~~\mbox{fermionic} \non\\
\ee
and consequently,
\be
&& (\pi \otimes \mbox{id}^{\otimes N}) \Delta^{'(N+1)}(x)\ {\mathbb T}(\lambda) =
{\mathbb T}(\lambda)\ (\pi \otimes \mbox{id}^{\otimes N}) \Delta^{'(N+1)}(x) \non\\
&& x \in U_q(gl({\mathrm m} -\alpha|{\mathrm n})) \otimes U_q(gl(\alpha)),
~~~\mbox{if} ~~\alpha ~~\mbox{bosonic} \non\\
&& x \in  U_q(gl({\mathrm m}|\tilde \alpha)) \otimes U_q(gl({\mathrm n}- \tilde \alpha))
~~~\mbox{if} ~~
\alpha = m + \tilde \alpha ~~\mbox{fermionic}. \non\\
\ee
Thus following the logic described in the case where  $K \propto {\mathbb I}$
we show that:
\be
&&\Big [t(\lambda),\ U_q(gl(\alpha|{\mathrm n})) \otimes U_q(gl({\mathrm m}-\alpha)) \Big ],
~~~\mbox{if} ~~\alpha ~~\mbox{bosonic} \non\\
&&\Big [ t(\lambda),\ U_q(gl({\mathrm m}|\tilde \alpha)) \otimes U_q(gl({\mathrm n}- \tilde \alpha)) \Big ]
~~~\mbox{if} ~~
\alpha = m + \tilde \alpha ~~\mbox{fermionic}.
\ee
It is clear that in the limit $q\to 1$ relevant isotropic results are recovered associated to the ${\cal Y}(gl({\mathrm m}|{\mathrm n}))$.

Consider the simplest non trivial cases, that is the $U_q(gl(2|1))$ and $U_q(gl(2|2))$ algebras with
diagonal $K$ matrices (\ref{kdiag}) with $\alpha =2$.  According to the preceding discussion the $U_q(gl(2|1))$ and
$U_q(gl(2|2))$
symmetries
break down to $U_q(gl(2)) \otimes u(1)$ and $U_q(gl(2)) \otimes U_q(gl(2))$ respectively. It is worth presenting the
associated
Casimir operators for the tow particular examples.
Consider first the $U_q(gl(2|1))$ case, then
from the $\lambda \to \pm \infty$ asymptotic behavior of the open transfer matrix we obtain:
\be
&& C^+ = q^2 \Delta^{(N)}(q^{\epsilon_1 +\epsilon_2})
\Big (q \Delta^{(N)}(q^{\epsilon_1 -\epsilon_2}) + q^{-1} \Delta^{(N)}(q^{-\epsilon_1 +\epsilon_2})+ (q-q^{-1})^2
\Delta^{(N)}(f_1)
\Delta^{(N)}(e_1)
\Big ) \non\\
&& C^- =q^2 \Delta^{(N)}(q^{-2 \epsilon_3}). \label{qcas0}
\ee
Notice that all $\epsilon_i, ~i =1,\ 2,\ 3$ elements commute with the transfer matrix and also the parenthesis in
the first line of (\ref{qcas0})
is essentially the typical  $U_q(sl_2)$ quadratic Casimir. $C^-$
is basically a $u(1))$ type quantity. It is thus clear that $C^+$ is the quadratic Casimir associated to $U_q(gl_2)$;
indeed in this case the
$U_q(gl(2|1))$ symmetry breaks down to
$U_q(gl(2)) \otimes u(1)$.
In general the implementation of boundary conditions described by the diagonal matrix (\ref{kdiag}) with $\alpha = {\mathrm m}$
obviously
breaks the
super symmetry to $U_q(gl({\mathrm m})) \otimes U_q(gl({\mathrm n}))$, so in fact the super algebra reduces to
two non
super symmetric
quantum algebras. Also, $C^+$ is the Casimir associated  $U_q(gl({\mathrm m}))$ to whereas
$C^-$ is the Casimir associated to $U_q(gl({\mathrm n}))$.

This will become more transparent when examining the $U_q(gl(2|2))$ case with boundary conditions described by
$K$ (\ref{kdiag}) and $\alpha =2$.
The associated Casimir operators are then given by
\be
&& C^+ = q^2 \Delta^{(N)}(q^{\epsilon_1 + \epsilon_2}) \Big ( q \Delta^{(N)}(q^{\epsilon_1 -\epsilon_2}) + q^{-1}
\Delta^{(N)}
(q^{-\epsilon_1 +\epsilon_2})+ (q-q^{-1})^2 \Delta^{(N)}(f_1)
\Delta^{(N)}(e_1) \Big )\non\\
&& C^- =  q^2 \Delta^{(N)}(q^{-\epsilon_3 - \epsilon_4}) \Big (q \Delta^{(N)}(q^{\epsilon_4 -\epsilon_3}) + q^{-1}
\Delta^{(N)}(q^{-\epsilon_4
+\epsilon_3})+ (q-q^{-1})^2 \Delta^{(N)}(f_3)
\Delta^{(N)}(e_3) \Big ). \non\\
\ee
As expected in this case, since now the symmetry is broken to $U_q(gl_2) \otimes U_q(gl_2)$, the $C^+$ Casimir
is associated to
the one $U_q(gl_2)$ symmetry whereas the $C^-$ is associated to the other $U_q(gl_2)$.

\subsubsection{Non-diagonal reflection matrices}

Let us finally consider non-diagonal reflection matrices.
A new class of non-diagonal reflection matrices associated to $U_q(gl{\mathrm m}|{\mathrm n})$
was recently derived in \cite{doikoukaraiskos}. Specifically,
first define the conjugate index $\bar a$ such that:
$[a] = [\bar a]$, and more specifically:
\be
&& \bar a = 2{\mathrm k} +{\mathrm m} +1 -a; ~~~~\mbox{Symmetric diagram} \non\\
&& \bar a = {\mathrm m} +1- a, ~~\mbox{$a$ bosonic}; ~~~\bar a = 2{\mathrm m}+{\mathrm n}+1-a,
~~\mbox{$a$ fermionic}; ~~~~
\mbox{Distinguished diagram.} \non\\
\ee
Then the non diagonal matrices read as:
\\
\\
{\bf Symmetric Dynkin diagram:}
\be
&& K_{aa}(\lambda) = e^{2 \lambda} \cosh i\mu m - \cosh 2 i \mu \zeta, ~~~K_{\bar a \bar a}(\lambda)=
e^{-2 \lambda} \cosh i\mu m -
\cosh 2i \mu \zeta,
\non\\ && K_{a \bar a}(\lambda)=
i c_a \sinh 2 \lambda, ~~~ K_{\bar a a}(\lambda)=
i c_{\bar a} \sinh 2 \lambda, ~~~~1\leq a\leq L
\non\\
&& K_{aa}(\lambda) = K_{\bar a \bar a}(\lambda)= \cosh (2\lambda +i m \mu) - \cosh 2i \mu \zeta, ~~~
K_{a\bar a}(\lambda)=  K_{\bar a a}(\lambda)=0, ~~~~L< a \leq {{\mathrm m} +2{\mathrm k} \over 2}; \non\\
&& 1 \leq L \leq {{\mathrm m} +2{\mathrm k} \over 2} \non\\
&& K_{AA} = \cosh (2\lambda +i m \mu) - \cosh 2i \mu \zeta, ~~~A={{\mathrm m}+2{\mathrm k} +1\over 2}~~~~
\mbox{if ${\mathrm m}$ odd}.
\label{k}
\ee
$m,\ \zeta$ are free boundary parameters.

Let us first focus on the solution with the minimal number of non-zero entries, and let us consider the symmetric case.
The elements ${\mathbb T}_{ij}^{\pm},\ i,\ j \in \{2, \ldots, 2{\mathrm k} + {\mathrm m} - 1\}$ form essentially
the $U_q(gl({\mathrm m}|2({\mathrm k}-1)))$
algebra, and it can
be shown, based on the logic described in the previous section, that the transfer matrix
commutes with  $U_q(gl({\mathrm m}|2({\mathrm k}-1)))$ plus the elements $ {\mathbb T}^{\pm}_{1j} \cup
{\mathbb T}^{\pm}_{j1} \cup {\mathbb T}^{\pm}_{{\cal N} j} \cup
{\mathbb T}^{\pm}_{j  {\cal N}}$ (${\cal N} = {\mathrm m} +2 {\mathrm k}$).
Let us now deal with the generic situation described by the solutions presented above (\ref{k}).
First define: ${\mathbb U} = {\mathbb T}^{\pm}_{Aj} \cup
{\mathbb T}^{\pm}_{jA} \cup {\mathbb T}^{\pm}_{\bar Aj} \cup
{\mathbb T}^{\pm}_{j \bar A}\ =0,
~~~A \in \{1,\ \ldots, L\}$.
Then it is shown for the generic non-diagonal $K$-matrix:
\\
\\
{\bf Symmetric Dynkin diagram}
\be
&& \Big [t(\lambda),\  U_q(gl({\mathrm m}|2({\mathrm k}-L)))\otimes {\mathbb U} \Big ] =0 ~~~\mbox{bosonic indices},
\non\\
&& \Big [t(\lambda),\  U_q(gl({\mathrm m-2l}))\otimes {\mathbb U} \Big ] =0 ~~~L={\mathrm k} +l,
 ~~~\mbox{fermionic indices}.
\ee
The generic Casimir is given by (\ref{qcas}), where now the only non zero entries are given by: $K_{aa},\ K_{a \bar a}$

It is clear that different choices of non-diagonal reflection matrices lead to distinct preserving symmetries.
This may perhaps be utilized
together with the contraction process presented in \cite{doikousfetsos} to offer an algebraic description regarding of the
underlying algebra emerging in the
AdS/CFT context.
Recall that this type of contractions leads to centrally extended algebras, and as in known in the context of AdS/CFT we deal
basically with a centrally extended $gl(2|2)$ algebra \cite{beisert}.

The explicit form of the boundary non local charges ${\mathbb T}{ab}$ (\ref{nonlocal})
in addition to the existence of some familiar symmetry is essential, and may be
for instance utilized for deriving reflection matrices associated to higher representations
of $U_q(gl({\mathrm m}|{\mathrm n}))$ (see e.g. \cite{nepomechiek}). In fact, the logic we follow here is
rather two-fold: on the one hand we try to extract a familiar symmetry algebra if any, and following
the process of
section 3.1.1 to derive the exact symmetry. On the other hand for generic $K$ that may break all familiar symmetries we show
via the reflection equation
that the boundary non-local charges form an algebra, and via (\ref{symme2}) we show that provide an extra symmetry for the open
transfer matrix.
Moreover,  the knowledge of the explicit form of the boundary non-local charges is of great significance given that
they may be used, as already mentioned, for
deriving reflection matrices for arbitrary representations \cite{nepomechiek}.

\subsection{The $q$ twisted super Yangian}

To complete our analysis on the boundary super symmetric algebras we shall now
briefly discuss the $q$
twisted Yangian. A more detailed analysis together with the classification of the corresponding
 $c$-number solutions will be pursued elsewhere.
As in the case of the twisted Yangian we focus on the symmetric Dynkin diagram (\ref{symm}) and
introduce
\be
V = \sum_i f_{i}\ e_{i \bar i}: ~~~~~V^T\ V =M
\ee
and define the super transposition as in the rational case. Also define the following matrices:
\be
\bar R_{12}(\lambda) := R_{21}^{t_1}(-\lambda - i \rho), ~~~~\bar R_{21}(\lambda) := R_{12}^{t_2}(-\lambda - i \rho).
\ee
Recall that in the isotropic case $R_{12} = R_{21}$.

Then the $q$-Twisted Yangian is defined by:
\be
R_{12}(\lambda_1-\lambda_2) K_1(\lambda_1) \bar R_{21}(\lambda_1 +\lambda_2) K_2(\lambda_2)=
K_2(\lambda_2) \bar R_{12}(\lambda_1+\lambda_2) K_1(\lambda_1) R_{21}(\lambda_1-\lambda_2).
\ee

The non-local charges are derived from the asymptotic behavior of the tensor representation of the $q$-twisted Yangian:
\be
{\mathbb T}^{\pm}_{ab} = \sum_{k} (-1)^{([a]+[k])([k]+[b])} T_{ak}^+ \hat T^+_{kb}.
\ee
As in the non super symmetric case the non local charges derived above satisfy exchange relations of the type
(see also e.g. \cite{crampedoikou})
\be
R^{\pm}_{12}\ {\mathbb T}^{\pm}_1\ \bar R^{\pm}_{21}\ {\mathbb T}^{\pm}_2=
{\mathbb T}^{\pm}_2\ \bar R^{\pm}_{12}\ {\mathbb T}^{\pm}_1\ R^{\pm}_{21}
\ee
it turns out that they
do {\it not} provide an exact symmetry of the
transfer matrix \cite{crampedoikou}, as opposed to the isotropic limit described in section 2.2.
And although in the context of discrete
integrable models described by the double row transfer matrix the boundary non-local charges
do not form an exact
symmetry one may show that in the corresponding field theoretical context they do provide exact symmetries
(see e.g. \cite{dema}).
Such investigations regarding super symmetric field theories will be left however for future investigations.
Note finally that the classification of solutions of the $q$ twisted Yangian for the
$U_q(gl({\mathrm m}| {\mathrm n}))$ is still
an open question, which we hope to address in a forthcoming publication.

\appendix

\section{Appendix}

We shall recall here some details regarding the $U_{q}(gl({\mathrm m}|{\mathrm n}))$ algebra.
The algebra is defined by generators $\epsilon_i$, $e_j\ f_j$, $i=1, \ldots , {\cal N},\ ~ j = 1, \dots ,
{\cal N} -1$,
and the exchange relations of the $U_{q}(gl({\mathrm m}|{\mathrm n}))$ algebra are given below:
\be
&& q^{\epsilon_i}\ q^{-\epsilon_i}=   q^{-\epsilon_i}\  q^{\epsilon_i} =1 \non\\
&& q^{\epsilon_i}\ e_j = q^{(-1)^{[j]}\delta_{ij} - (-1)^{[j+1]}\delta_{i j+1}}e_j\ q^{\epsilon_i}\non\\
&& q^{\epsilon_i}\ f_j = q^{-(-1)^{[j]}\delta_{ij} + (-1)^{[j+1]}\delta_{i j+1}}f_j\ q^{\epsilon_i}\non\\
&& e_i f_j - (-1)^{([i]+[i+1])([j]+[j+1])}f_j e_i = \delta_{ij} {q^{\epsilon_i - \epsilon_{i+1}}-
q^{-\epsilon_i + \epsilon_{i+1}} \over q - q^{-1}}\non\\
&& x_i\ x_j= (-1)^{([i]+[i+1])([j]+[j+1])} x_j\ x_i, ~~~~x_i \in \{e_i,\ f_i\},
\ee
and $q$ Chevallay-Serre relations:
\be
x_i^2x_{i \pm 1} - (q+q^{-1})x_i x_{i \pm 1}x_i + x_{i\pm 1}x_i^2 =0, ~~~~x_i \in \{e_i,\ f_i\}~~~~
i \neq {\mathrm m}.
\label{cs}
\ee
Now set $h_i = \epsilon_i $ then the $U_q(sl({\mathrm m}|{\mathrm n}))$  algebra is defined\ by generators
$e_i,\ f_i,\ h_i$. Let $a_{ij}$
the elements
of the
related Cartan ${\cal N} \times {\cal N}$ matrix, which for instance for the distinguished Dynkin diagram is:
\be a = \left(\begin{array}{ccccccc}
2 &-1 &  &\ & & &\\
-1 &2 &-1 & & & & \\
 &  &\ddots &   &  &   &\\
 &&-1 &0 &1 & &\\
 &  &  &   &\ddots   &   &\\
 &  &   &  &1  & -2 &1\\
 & &  &  &  &1 &-2
\end{array}\right) \label{tr2} \ee
the zero diagonal element occurs in the ${\mathrm m}$ position.
Define also:
\be
[h]_q= {q^{h} -q^{-h} \over q- q^{-1} }
\ee
then the $U_q(sl({\mathrm m}|{\mathrm n}))$ super algebra for the distinguished Dynkin diagram reads as:
\be
&& [e_i, f_i]= [h_i]_q, ~~~~i< {\mathrm m}, ~~~\{e_{\mathrm m},\ f_{\mathrm m}\}= - [h_{\mathrm m}]_q,
~~~~[e_i, f_i]= - [h_i]_q, ~~~~i> {\mathrm m}
\non\\
&& [h_j,\ h_k]=0, ~~~[h_i,\ e_j] = a_{ij}e_j, ~~~~[h_i,\ f_j]= -a_{ij} f_j
\ee
plus the Chevalley-Serre relations (\ref{cs}).
The algebra above is equipped with a non-trivial co-product:
\be
&& \Delta(\epsilon_i) = \epsilon_i \otimes {\mathbb I} + {\mathbb I} \otimes \epsilon_i \non\\
&& \Delta(x) = q^{-{h_i \over 2}} \otimes x + x \otimes q^{{h_i \over 2}}, ~~~~~x\in\{e_i,\ f_i\}.
\ee

There is also an isomorphism between the FRT representation of the algebra and the Chevalley-Serre basis.
Recall that
\be
&& L(\lambda)= L^+ - e^{-2\lambda} L^-, \non\\
&& \hat L(\lambda)= \hat L^+  - {\cal O}(e^{-2 \lambda}), ~~~\lambda \to \infty \non\\
&& \hat L(\lambda)= \hat L^-  - {\cal O}(e^{2 \lambda}), ~~~\lambda \to -\infty.
\ee
Recall for the reflection
algebra
$\hat L(\lambda) = L^{-1}(-\lambda)$ and also
\be
L^{\pm } =\sum_{i, j} e_{ij} \otimes l^{\pm}_{ij},\ ~~~~ \hat L^{\pm } =\sum_{i, j} e_{ij} \otimes \hat
l^{\pm}_{ij}.
\ee
Then we have the following identifications \cite{tak, difre}:
\be
&& l^+_{ii} = (-1)^{[i]} q^{\epsilon_i}, ~~~~l^+_{i i+1} = (-1)^{[i+1]} (q -q^{-1}) q^{{\epsilon_i +
\epsilon_{i+1}\over 2}}f_i,
~~~l_{i+1 i}^+ =0 \non\\
&& l^-_{ii} = (-1)^{[i]}q^{-\epsilon_i}, ~~~~l^-_{i+1 i} = -(-1)^{[i]} (q -q^{-1})e_i  q^{-{\epsilon_i +
\epsilon_{i+1}\over 2}},
~~~l^-_{i i+1} =0\non\\
&& \hat l^+_{ii} = (-1)^{[i]} q^{\epsilon_i}, ~~~~\hat l^+_{i+1 i} = (-1)^{[i+1]} (q -q^{-1})q^{-{a_{ii}\over 2}}
q^{{\epsilon_i + \epsilon_{i+1}\over 2}}e_i,
~~~\hat l_{i i+1}^+ =0 \non\\
&& \hat l^-_{ii} = (-1)^{[i]}q^{-\epsilon_i}, ~~~~\hat l^-_{i+1 i} = -(-1)^{[i]} (q -q^{-1}) q^{{a_{ii}\over 2}}f_i
q^{-{\epsilon_i + \epsilon_{i+1}\over 2}},
~~~\hat l^-_{ i+1 i} =0
\ee
$l_{ij}^+,\ \hat l^-_{ij}, ~~i<j$ and $l_{ij}^-,\ \hat l^+_{ij}, ~~i>j$ are also non zero, and are expressed as
combinations
of the $U_q(gl({\mathrm m}|{\mathrm n}))$ generators, but are omitted here for brevity, see e.g expressions in
\cite{jimbo, myhecke} for the $U_q(gl({\mathrm n}))$ case.

\end{document}